\documentclass
[aps,preprint,showkeys,a4paper,tightenlines,superscriptaddress]{revtex4}%
\usepackage{eurosym}
\usepackage{amsfonts}
\usepackage{amsmath}
\usepackage{amssymb}
\usepackage{graphicx}
\usepackage{subfig}
\usepackage{caption}%
\providecommand{\U}[1]{\protect\rule{.1in}{.1in}}
\providecommand{\U}[1]{\protect\rule{.1in}{.1in}}
\newtheorem{theorem}{Theorem}
\newtheorem{acknowledgement}[theorem]{Acknowledgement}

\begin{document}
\title{Exploring the effect of geometric coupling on friction and energy dissipation
in rough contacts of elastic and viscoelastic coatings}
\author{N. Menga}
\email{nicola.menga@poliba.it}
\affiliation{Department of Mechanics, Mathematics and Management, Politecnico of Bari, V.le
Japigia, 182, 70126, Bari, Italy}
\affiliation{Imperial College London, Department of Mechanical Engineering, Exhibition
Road, London SW7 2AZ}
\author{G. Carbone}
\affiliation{Department of Mechanics, Mathematics and Management, Politecnico of Bari, V.le
Japigia, 182, 70126, Bari, Italy}
\affiliation{Imperial College London, Department of Mechanical Engineering, Exhibition
Road, London SW7 2AZ}
\author{D. Dini}
\affiliation{Imperial College London, Department of Mechanical Engineering, Exhibition
Road, London SW7 2AZ}
\begin{abstract}
We study the frictional behavior of both elastic and viscoelastic thin
coatings bonded to a seemingly rigid substrate and sliding against a rough
profile in the presence of Coulomb friction at the interface. The aim is to
explore the effect of the coupling between the normal and tangential
displacement fields arising from the finiteness of the material thickness and
to quantify the contribution this can have on energy losses.

We found that, due to normal-tangential coupling, asymmetric contacts and
consequently additional friction are observed even for purely elastic layers,
indeed associated with zero bulk energy dissipation. Furthermore, enhanced
viscoelastic friction is reported in the case of viscoelastic coatings due to
coupling, this time also entailing larger bulk energy dissipation.

Geometric coupling also introduces additional interactions involving the
larger scales normal displacements, which leads to a significant increase of
the contact area, under given normal load, compared to the uncoupled contacts.

These results show that, in the case of contact interfaces involving thin
deformable coating bonded to significantly stiffer substrate, the effect of
interfacial shear stresses on the frictional and contact behavior cannot be neglected.

\end{abstract}
\maketitle

\section{Introduction}

Nowadays, a large number of systems in several application fields involve thin
solid films. Soft coatings are one of the most frequent examples of such
cases: a thin layer of compliant material with specific characteristics is
deposited onto a significantly stiffer substrate (thus considered as rigid) in
order to tailor the overall system behavior (e.g. chemical resistance to
corrosion, enhanced or reduced stiffness, damping, frictional behavior).
Possible applications range from engines, where specific low friction coatings
are adopted to reduce energy dissipation in key contacting pairs (e.g. valve
train systems and crankshafts \cite{Dahotre2005,Kano2006}), to robotic clamps
for objects manipulation \cite{Voigt2012} or anti-skid tapes for ramps and
stairs, where high frictional coatings are instead required to increase the
grip. Coatings are also present in the case of biological systems such as
human hands and feet, where the covering skin (which may locally be
constituted by very thin layers) concurs in developing the high interfacial
friction sustaining, for instance, the firm hand grip on the tennis racket
handle, or the barefoot walking on different grounds.

For these reasons, a constantly rising interest on the tribological behavior
of solid thin films, often studied as compliant layers of materials bonded to
rigid bodies and indented by other rigid or deformable rough contersurfaces,
has been reported in the last decades. Indeed, besides the theoretical
\cite{GW1966,BGT,Persson2001,YangPersson2008,Menga2014,menga2018,menga2018corr,menga2018bis}%
, numerical \cite{Hyun2004, Campana2008,
Pastewka2016,DiniMedina,Muser2017,mengaVpeeling} and experimental
\cite{Homola1990,Chateauminois2008,new2,Fineberg2010} studies focusing on
contact problems of semi-infinite bodies, detailed investigations have been
also devoted to the case of contacts involving thin bodies
\cite{Carbone2008,Putignano2015,menga2016,menga2016visco,menga2018visco,mengaRLRB,mengaVpeelingthin}%
.

To this regard, it is well known that dealing with half-space contacts, a
certain degree of coupling between the normal and tangential displacement
fields occurs in the case of material dissimilarity
\cite{HillsBook,BarberBook}. Such a \textit{material} coupling is governed by
the Dundurs' second constant, often referred to as $\beta$, which if one of
the bodies is rigid takes the value $\beta=(1-2\nu)/2(1-\nu),$ with $\nu$
being the Poisson's ratio. This effect has been explored in several studies,
mostly focusing on stick-slip fretting problems associated to homogeneous
\cite{Nowell1988,Chen2008,Chen2009} and graded \cite{Wang2010,Elloumi2010}
elastic materials. Interestingly, in \cite{Nowell1988} it was reported that,
in the case of dissimilar cylinders contacts, a non-negligible influence of
the material coupling occurs on both the normal stiffness of the contact and
the contact pressure distribution. However, a few pioneeristic studies
\cite{Bentall1968,Nowell1988b,Nowell1988c}, dealing with thin deformable
layers, have shown that such a simple coupling representation is no longer
valid as, since the normal deformation cannot be accommodated remotely as in
the case of half-space, two possible independent sources of normal-tangential
interactions exist: (i) the \textit{material} coupling, due to material
dissimilarity, governed by the Dundurs' second constant $\beta$; and (ii) an
additional \textit{geometric} (or domain shape) coupling, which depends on the
layer thickness (i.e. it vanishes for thick layers) and still occurs even in
the case of similar contact pairs (i.e. $\beta=0$). Significantly less effort
has been made to investigate the effect of the latter on the contact behavior
of thin films. Indeed, moving from the pioneeristic study of Bentall and
Johnson \cite{Bentall1968}, only a few authors have approached the problem
\cite{Nowell1988b,Nowell1988c,Jaffar1993,Jaffar1997} focusing on smooth single
asperity contacts and showing a significant contact pressure asymmetry arising
from the coupling. Furthermore, in a recent study \cite{menga2019geom}, the
rough contact behavior of elastic thin layers in the presence of interfacial
friction has been investigated, showing that, even in the case of $\beta=0$,
the \textit{geometric} coupling between the normal and tangential elastic
fields may lead to a significant increase of the effective contact area, with
non-negligible implication on contact-related phenomena such as interfacial
hydraulic impedance, electrical conductivity \cite{Kogut2003}, and wear
process evolution \cite{MengaCiava}. Interestingly, the \textit{geometric}
coupling may play an even more dramatic role in determining the frictional
performance of interfaces in relative motion due to the asymmetry of the
contact pressure distribution observed in Refs.
\cite{Nowell1988b,Nowell1988c,menga2019geom}. Focusing, for instance, on
viscoelastic contact of thin layers, one can reasonably expect different
energy dissipation due to bulk viscoelasticity and, in turn different
frictional behavior of the interface, depending on the specific
\textit{geometric} coupling effect on the contact pressure and contact spots
distribution which alter the effective excitation spectra during sliding. To
the best of the authors knowledge, an investigation on this effect is
currently missing in the specialized literature, and this work aims at filling
this gap.

In this study we focus on the case of a thin coating, sufficiently softer than
the underlying substrate so that the latter can be assumed as rigid, in
frictional sliding contact with a rigid profile with self-affine roughness. We
consider both elastic and viscoelastic coating materials. We investigate in
details the effect of normal-tangential coupling in thin films on both the
overall contact behavior and frictional response of the system, with further
focus on the energy dissipation. As already mentioned, the system
configuration studied here covers several technological applications related
to the grip performance of bio-inspired or natural system for handling of
objects as well as many other interesting problems, including protein-coated
interfaces, paints and soft coatings for industrial use, finger tip contact
with touch screens.

\section{The contact problem formulation}

\begin{figure}[ptbh]
\begin{center}
\includegraphics[width=0.8\textwidth]{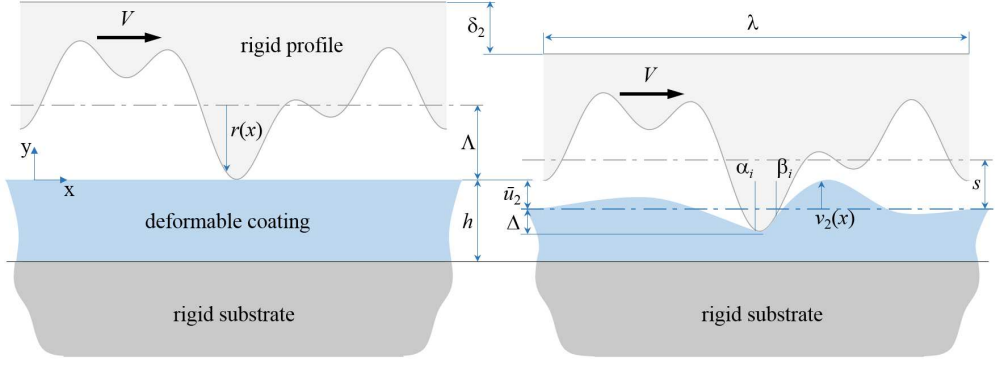}
\end{center}
\caption{A soft coating of thickness $h$ backed onto a rigid substrate is in
sliding contact with a rigid rough profile. Coulomb friction interactions
occur at the interface. The contact mean penetration is indicated with
$\Delta$, the profile peak height is $\Lambda$, and the roughness fundamental
wavelength is $\lambda$.}%
\label{fig1}%
\end{figure}

The system under investigation is shown in Figure \ref{fig1}, where a thin
soft coating bonded to a rigid substrate is sketched. The free surface of the
coating layer is indented by a rigid profile with roughness $r\left(
x\right)  $. According to Fig. \ref{fig1}, we define $h$ the coating
thickness, $\lambda$ the roughness fundamental wavelength, and $V$ the profile
sliding speed. In our formulation, we assume $V\ll c_{s}$ with $c_{s}$ being
the sound speed into the coating material; furthermore, we focus on long time
observations so that steady state conditions can be reasonably assumed. In
what follows, we will adopt subscript $1$ and $2$ referring to tangential and
normal quantities, respectively. Indeed, in Fig. \ref{fig1}, $\delta_{2}$ is
the total normal displacements of the rough profile, $\bar{u}_{2}$ is the mean
normal displacement of the coating surface, whereas $\Delta$ is the mean
penetration of the rigid profile into the deformable coating. Note that
$\delta_{2}=\Delta+\bar{u}_{2}$.

The presence of Amonton/Coulomb friction is taken into account at the contact
interface. This means that, given a generic normal contact pressure
distribution $p\left(  x\right)  $, a corresponding tangential shear stress
distribution also acts on the contacting parts in the form
\begin{equation}
\tau\left(  x\right)  =\mu_{c}p\left(  x\right)  ;\ \ \ \ \ x\in\Omega,
\label{frict}%
\end{equation}
where $\mu_{c}$ is the friction coefficient, and $\Omega=U_{i=1}^{L}\left[
\alpha_{i},\beta_{i}\right]  $ is the contact domain, being $\alpha_{i}$ and
$\beta_{i}$ the unknown coordinates of \textit{i-}th contact spot, with
$\alpha_{i}<\beta_{i}$ and $i=1,2,...,L$, where $L$ is the unknown number of
contacts. Notably, we assume $\mu_{c}$ independent of the relative sliding speed.

The contact problem approach exploits the reliable formulation developed in
Refs \cite{carb-mang-2008,menga2018visco}. Indeed, building on the linearity
of the material response and exploiting the problem translational invariance,
the interfacial layer displacement vector $\mathbf{v}=\left(  v_{1}%
,v_{2}\right)  $ can be linked to the stress vector $\mathbf{\sigma}=\left(
\mu_{c}p,-p\right)  $ by means of%
\begin{equation}
\mathbf{v}\left(  x\right)  =\mathbf{u}\left(  x\right)  -\mathbf{\bar{u}%
}=\int_{\Omega}ds\Theta\left(  x-s\right)  \mathbf{\sigma}\left(  s\right)
;\text{ \ \ \ \ \ }x\in\Omega,\label{10}%
\end{equation}
where the time-dependency of the stress and deformation fields has been
removed by invoking the coordinate transformation $x-Vt\rightarrow x$.

In Eq. (\ref{10}), $\mathbf{u}$ is the total displacement vector, and
$\mathbf{\bar{u}}$ is the mean displacement vector given by%
\begin{align}
&  \bar{u}_{\mathrm{1}}=q_{0}h\frac{1+\nu}{\pi E_{0}}\tau_{\mathrm{m}}%
\lambda,\label{um1}\\
&  \bar{u}_{\mathrm{2}}=q_{0}h\frac{1-2\nu}{2\pi E_{0}}p_{\mathrm{m}}\lambda,
\label{um2}%
\end{align}
where $q_{0}=2\pi/\lambda$ and $E_{0}$ is zero-frequency elastic modulus.
Notably, $p_{m}=\frac{1}{\lambda}\int_{\Omega}p\left(  x\right)  dx$ and
$\tau_{m}=\frac{1}{\lambda}\int_{\Omega}\tau\left(  x\right)  dx$ are the mean
contact pressure and shear stress, respectively.

The term $\Theta\left(  x\right)  =\Theta_{kl}\left(  x\right)  $, with $k,l$
$=1,2$ is the Green's tensor which takes different forms depending on whether
the coating is elastic or viscoelastic. Indeed, in the elastic case, we have
\begin{equation}
\Theta_{kl}\left(  x\right)  =\frac{G_{kl}\left(  x\right)  }{E_{0}},
\end{equation}
with $G_{kl}\left(  x\right)  $ given by Ref. \cite{menga2019geom} as
\begin{align}
G_{11}\left(  x\right)   &  =-\frac{2\left(  1-\nu^{2}\right)  }{\pi}\left[
\log\left\vert 2\sin\left(  \frac{q_{0}x}{2}\right)  \right\vert +\sum
_{m=1}^{\infty}B\left(  mq_{0}h\right)  \frac{\cos\left(  mq_{0}x\right)  }%
{m}\right]  ,\label{G11}\\
G_{12}\left(  x\right)   &  =-G_{21}\left(  x\right)  =\frac{1+\nu}{\pi
}\left[  \frac{1-2\nu}{2}\left[  \text{sgn}\left(  x\right)  \pi
-q_{0}x\right]  -\sum_{m=1}^{\infty}C\left(  mq_{0}h\right)  \frac{\sin\left(
mq_{0}x\right)  }{m}\right]  ,\label{G12}\\
G_{22}\left(  x\right)   &  =-\frac{2\left(  1-\nu^{2}\right)  }{\pi}\left[
\log\left\vert 2\sin\left(  \frac{q_{0}x}{2}\right)  \right\vert +\sum
_{m=1}^{\infty}A\left(  mq_{0}h\right)  \frac{\cos\left(  mq_{0}x\right)  }%
{m}\right]  , \label{G22}%
\end{align}
with
\begin{align}
A\left(  mq_{0}h\right)   &  =1+\frac{2mq_{0}h-\left(  3-4\nu\right)
\sinh\left(  2mq_{0}h\right)  }{5+2\left(  mq_{0}h\right)  ^{2}-4\nu\left(
3-2\nu\right)  +\left(  3-4\nu\right)  \cosh\left(  2mq_{0}h\right)
},\label{A}\\
B\left(  mq_{0}h\right)   &  =1-\frac{2mq_{0}h+\left(  3-4\nu\right)
\sinh\left(  2mq_{0}h\right)  }{5+2\left(  mq_{0}h\right)  ^{2}-4\nu\left(
3-2\nu\right)  +\left(  3-4\nu\right)  \cosh\left(  2mq_{0}h\right)
},\label{B}\\
C\left(  mq_{0}h\right)   &  =\frac{4\left(  1-\nu\right)  \left[  2+\left(
mq_{0}h\right)  ^{2}-6\nu+4\nu^{2}\right]  }{5+2\left(  mq_{0}h\right)
^{2}-4\nu\left(  3-2\nu\right)  +\left(  3-4\nu\right)  \cosh\left(
2mq_{0}h\right)  }. \label{C}%
\end{align}

On the other hand, in the case of viscoelastic coatings, extending the
formulation given in Refs. \cite{menga2016visco,menga2018visco}, the Green's
tensor takes the form%

\begin{equation}
\Theta_{kl}^{V}\left(  x\right)  =J\left(  0^{+}\right)  G_{kl}\left(
x\right)  +\int_{0^{+}}^{+\infty}G_{kl}\left(  x+Vt\right)  \dot{J}\left(
t\right)  dt, \label{eq_visc}%
\end{equation}
which, this time, in order to take into account for the response delay in the
viscoelastic material, parametrically depends on the sliding speed. In Eq.
(\ref{eq_visc}), the tern $J\left(  t\right)  $ is the viscoelastic creep
function, which for a linear viscoelastic material with one relaxation time
$\tau$ takes the form%
\begin{equation}
J\left(  t\right)  =H\left(  t\right)  \left[  \frac{1}{E_{0}}-\frac{1}{E_{1}%
}\exp\left(  -t/\tau\right)  \right]  , \label{3}%
\end{equation}
where $H\left(  t\right)  $ is the Heavyside step function, and $1/E_{1}%
=1/E_{0}-1/E_{\infty}$, with $E_{\infty}$ being the high-frequency elastic modulus.

Regardless of the coating material rheology, the solution of the contact
problem is found by observing that, within the contact domain $\Omega$, the
normal interfacial displacement must match the rough profile shape $r\left(
x\right)  $. Referring to Fig. \ref{fig1}, from the normal projection of Eq.
(\ref{10}) we have%
\begin{equation}
\Lambda-r\left(  x\right)  -\Delta=\int_{\Omega}ds\left[  \mu_{c}\Theta
_{21}\left(  x-s\right)  -\Theta_{22}\left(  x-s\right)  \right]  p\left(
s\right)  ;\text{ \ \ \ \ \ }x\in\Omega, \label{10n}%
\end{equation}
where the only unknowns are the pressure distribution $p\left(  x\right)  $
and the coordinates $\alpha_{i}$, $\beta_{i}$ corresponding to the individual
contact edges. By relying on the numerical strategy based on a non-uniform
contact area discretization developed in Ref. \cite{carb-mang-2008}, for any
given value of the contact penetration $\Delta$, Eq. (\ref{10n}) can be
numerically solved for $p\left(  x\right)  $,$~$once fixed $\alpha_{i}$,
$\beta_{i}$. Further, the exact size of the contact area can be calculated,
iteratively, by observing that, dealing with adhesiveless contact conditions,
bounded contact stress are prescribed, specifically vanishing at the contact
edges. Indeed, referring to Refs. \cite{maugis,menga2016visco,menga2018visco}
we have that%
\begin{align}
K_{I,\alpha_{i}}  &  =-\lim_{x\rightarrow\alpha_{i}^{+}}\sqrt{2\pi\left(
x-\alpha_{i}\right)  }p\left(  x\right)  =0,\\
K_{I,\beta_{i}}  &  =-\lim_{x\rightarrow\beta_{i}^{-}}\sqrt{2\pi\left(
\beta_{i}-x\right)  }p\left(  x\right)  =0,
\end{align}
where $K_{I}$ is the mode I stress intensity factor.

\section{Results and Discussion}

The presence of a deformable layer of finite thickness gives rise to coupling
between the normal and tangential displacement fields
\cite{HillsBook,BarberBook}. Indeed, in agreement with
\cite{Bentall1968,menga2019geom}, focusing on the cross-coupled Green's
function $G_{12}$ given in Eq. (\ref{G12}), we observe two coupling terms: the
first right-hand side term represents\ the \textit{material} coupling, taking
into account for the normal-tangential interactions in contact pairs of
dissimilar materials. Notably, by recalling the Dundurs' second constant
expression assuming one of the contacting bodies as rigid $\beta
=(1-2\nu)/2(1-\nu)$, we observe that material coupling vanishes for $\beta=0$
\cite{HillsBook,BarberBook,Nowell1988}. The second right-hand side term in Eq.
(\ref{G12}) is an additional source of normal-tangential\textit{ }coupling,
this time called \textit{geometric}, which is a function of the layer size
through the thickness $h$, and is non-vanishing even for $\beta=0$. Notably,
in the limit of semi-infinite bodies (i.e. for $h\rightarrow\infty$) the
latter term vanishes, thus leading to the well-known half-plane or half-space
behavior for which normal-tangential coupling only depends on the value of
$\beta$ \cite{HillsBook,BarberBook}.

In what follows, we focus on the case of $\beta=0$ (i.e. $\nu=0.5$) so that
the whole normal-tangential coupling arises from the finiteness of the layer
thickness $h$ (the case with $\beta\neq0$ is briefly discussed in Appendix).
We aim at investigate the frictional behavior of both elastic and viscoelastic
interfaces in coupled conditions, with specific focus on the interfacial and
bulk energy dissipation. Furthermore, the contact behavior of purely
viscoelastic rubber-like coatings is also investigated in terms of mean
contact quantities (i.e. contact area size, contact mean pressure and
penetration, displacement field). In order to highlight the specific effect of
geometric normal-tangential coupling on the investigated quantities we compare
the results against those related to frictionless conditions (i.e. $\mu_{c}%
=0$), where no coupling effect occurs\cite{Nowell1988}.

All the calculations have been performed considering a self-affine roughness
on the rigid profile. The different profile shapes have been numerically
generated by exploiting the technique reported in Ref. \cite{menga2018visco}.
The Power Spectral Density (PSD) $C_{r}\left(  q\right)  =\left(  2\pi\right)
^{-1}\int dx\left\langle r\left(  0\right)  r\left(  x\right)  \right\rangle
e^{-iqx}$ of the considered roughness is given by%
\begin{align}
C_{r}\left(  q\right)   &  =C_{0}\left(  \frac{\left\vert q\right\vert }%
{q_{0}}\right)  ^{-\left(  2H+1\right)  };\text{ \ \ \ \ \ }q\in\left[
q_{0},q_{1}\right] \nonumber\\
C_{r}\left(  q\right)   &  =0;\text{ \ \ \ \ \ }q\notin\left[  q_{0}%
,q_{1}\right]  \label{roughPSD}%
\end{align}
where $q_{1}=Nq_{0}$ (being $N$ the number of roughness scales) and $H$ is the
Hurst exponent, which is related to the fractal dimension by $D_{f}=2-H$.
Profiles are generated assuming a root mean square roughness of $r_{rms}%
=\sqrt{\left\langle r^{2}\right\rangle }=10$ $\mu m$, $H=0.7$, and $N=100$.
Notably, since $q_{0}=2\pi/\lambda$, adjusting the value of $\lambda$ also
modifies the profile average square slope $m_{2}=\left\langle r^{\prime
2}\right\rangle =\int q^{2}C_{r}\left(  q\right)  dq$. Moreover, in order to
obtain a statistically significant contact behavior, results have been
ensemble averaged on several realizations for each value of the contact
parameter investigated, and are shown in terms of the following dimensionless
quantities: $\tilde{h}=q_{0}h$, $\tilde{a}=q_{0}a$, $\tilde{\Delta}%
=\Delta/\Lambda$, $\tilde{v}=v/\Lambda$, $\tilde{\Lambda}=q_{0}\Lambda$,
$\zeta=V\tau q_{0},$ and $\tilde{p}=2\left(  1-\nu^{2}\right)  p/\left(
E_{0}q_{0}\Lambda\right)  $. In viscoelastic calculations we assume
$E_{\infty}/E_{0}=3$.

\subsection{Frictional behavior}

In this paper we consider Coulomb friction interactions, which occurs through
distributed tangential tractions at the interface, as indicated in Eq.
(\ref{frict}). However, in the presence of asymmetric distributions of contact
pressure and normal displacement such as those resulting from viscoelastic
relaxation delay in sliding contact or normal-tangential anti-symmetric
coupling, an additional term on tangential force opposing the motion arises as
the tangential component of the normal pressure distribution projected along
the local rough profile normal direction. This friction force $F_{a}$, induced
by asymmetric pressure distribution, is usually calculated as
\cite{Persson2001,menga2018visco}
\begin{equation}
F_{a}=\int_{L}p\left(  x\right)  v_{2}^{\prime}\left(  x\right)  dx\label{Fa}%
\end{equation}
where $v_{2}^{\prime}$ is the first spatial derivative of the normal
displacement field, and $L=n\lambda$ is the rigid profile length. The
corresponding friction coefficient is%
\begin{equation}
\mu_{a}=\frac{F_{a}}{Lp_{m}},\label{mua}%
\end{equation}
so that the total friction force $F_{t}$ opposing the motion can then be
written as
\begin{equation}
F_{t}=\left(  \mu_{c}+\mu_{a}\right)  Lp_{m}\label{Ft}%
\end{equation}

Moreover, since several studies \cite{Persson2001,menga2016visco} have shown
that, under given contact area size, $\mu_{a}\propto\sqrt{m_{2}}$, here we
normalize friction results by the factor $\sqrt{m_{2}}$.

\subsubsection{Elastic contacts}

Elastic rough contacts are usually not affected by friction force $F_{a}$, as
for elastic contacts involving uncoupled half-space (i.e. $h\rightarrow\infty$
and $\beta=0$ so that $G_{12}=0$) symmetric pressures and displacements are
expected. However, when dealing with sufficiently thin films, even for similar
contact pairs material (i.e. $\beta=0$) \textit{geometric} coupling occurs.
Since from Eq. (\ref{G12}) the resulting normal-tangential coupling function
is an odd function of $x$, its effect is to introduce a certain degree of
asymmetry in the contact pressure and normal displacement distributions. The
resulting frictional force\ $F_{a}$\ opposing the motion can be calculated
from Eq. (\ref{Fa}) in the Fourier domain as
\begin{equation}
F_{a}=\frac{\mu_{c}E_{0}}{2\pi}\int dq\frac{S_{12}\left(  \left\vert
q\right\vert h\right)  }{S_{22}^{2}\left(  \left\vert q\right\vert h\right)
+\mu_{c}^{2}S_{12}^{2}\left(  \left\vert q\right\vert h\right)  }%
q^{2}\left\vert v_{2}\left(  q\right)  \right\vert ^{2}\label{Fa_el}%
\end{equation}
where $v_{2}\left(  q\right)  =\int dxv\left(  x\right)  e^{-iqx}$ is the
Fourier transform of the normal displacement field, and $S_{11}\left(
\left\vert q\right\vert h\right)  =2\left(  1-\nu^{2}\right)  \left[
1-B\left(  qh\right)  \right]  $, $S_{12}\left(  \left\vert q\right\vert
h\right)  =\left(  1+\nu\right)  \left[  C\left(  qh\right)  -\left(
1-2\nu\right)  \right]  $, $S_{22}\left(  \left\vert q\right\vert h\right)
=2\left(  1-\nu^{2}\right)  \left[  1-A\left(  \left\vert q\right\vert
h\right)  \right]  $. Notably, $F_{a}=0$ in the case of both frictionless
(i.e. $\mu_{c}=0$) or uncoupled (i.e. $h\rightarrow\infty$ and $\beta=0$) contacts.

It is interesting to observe that, since the layer material is purely elastic,
such the friction force $F_{a}$ arising from asymmetric contact pressure
distribution does not involve bulk energy dissipation. Indeed, by invoking the
energy balance applied to the deformable coating we have that%
\begin{equation}
V\left[  \int_{\Omega}p\left(  x\right)  v_{2}^{\prime}\left(  x\right)
dx-\mu_{c}\int_{\Omega}p\left(  x\right)  v_{1}^{\prime}\left(  x\right)
dx\right]  =0 \label{energ}%
\end{equation}
where $v_{1}^{\prime}$ is the first spatial derivative of the tangential
displacement field.

\begin{figure}[ptbh]
\begin{center}
\centering\subfloat[\label{fig_elast3}]{\includegraphics[height=49mm] {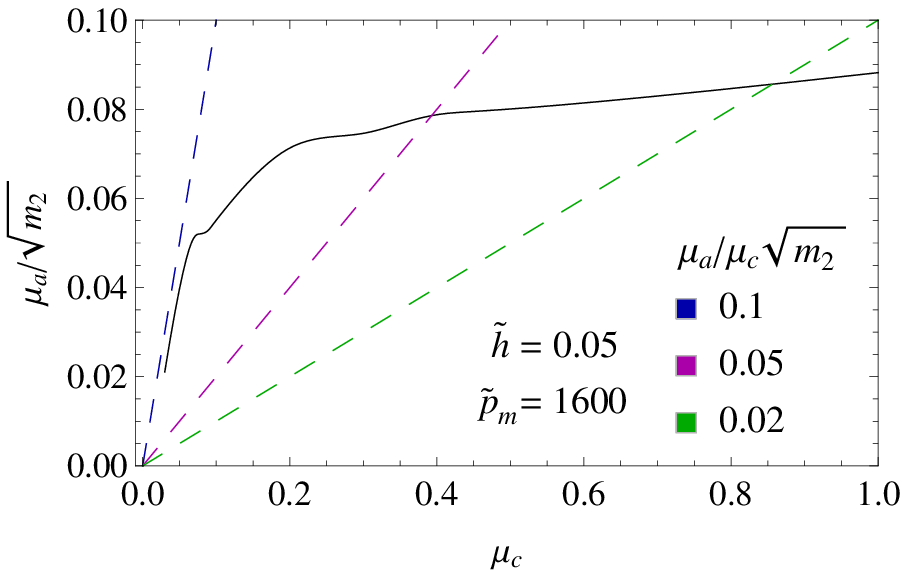}}
\hfill
\subfloat[\label{fig_elast4}]{\includegraphics[height=49mm]{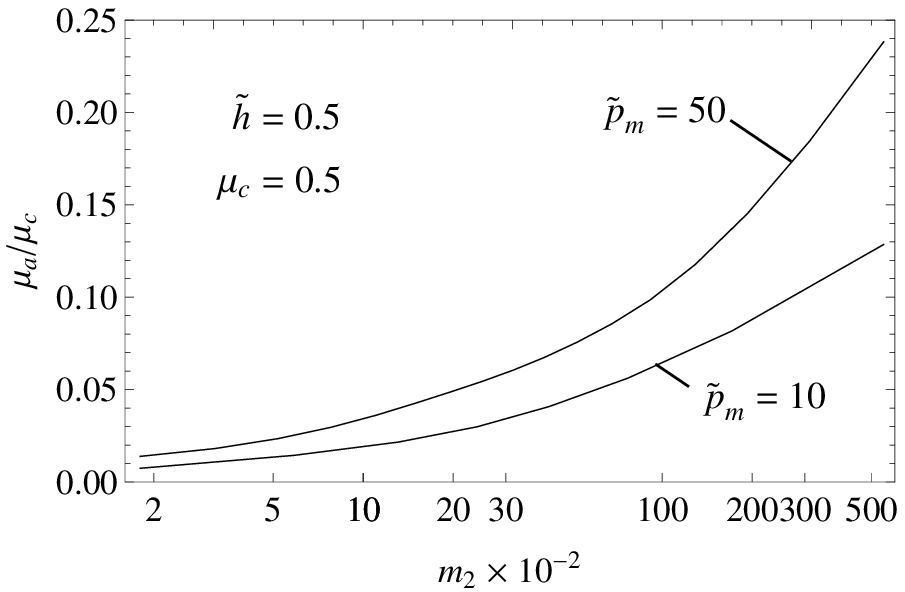}}
\end{center}
\caption{The normalized friction coefficient $\mu_{a}/\sqrt{m_{2}}$ due to
asymmetric coupled contact pressure as a function of the interfacial Coulomb
friction coefficient $\mu_{c}$ (a). The friction ratio $\mu_{a}/\mu_{c}$ as a
function of the rigid profile mean square slope $m_{2}$ (b). All the data
refer to $\beta=0$, and $m_{2}=0.018$ for (a).}%
\label{fig_elast}%
\end{figure}

Figures \ref{fig_elast} show the behavior of the friction coefficient $\mu
_{a}$ in elastic contacts of thin layers with $\beta=0$. Specifically, Figure
\ref{fig_elast1} reports the value of the normalized friction coefficient
$\mu_{a}/\sqrt{m_{2}}$ due to pressure asymmetry as a function of the
interfacial Coulomb friction coefficient $\mu_{c}$ for a given layer
thickness. We observe that, in agreement with Eq. (\ref{Fa_el}), under load
controlled conditions, reducing the tangential stresses (i.e. reducing
$\mu_{c}$) leads to an overall reduction of $\mu_{a}$. Indeed, the overall
normal-tangential coupling is modulated by Coulomb interfacial friction from
(see Eq. (\ref{10})), therefore reducing $\mu_{c}$ results in a lower degree
of asymmetry of the contact pressure distribution, and in turn of $\mu_{a}$.
Moreover, Figure \ref{fig_elast1} also investigates the friction ratio
$\mu_{a}/\mu_{c}$. This quantity clearly highlight the impact of coupling
effect on the overall friction opposing the indenter motion (see Eq.
(\ref{Ft})). To this regard, Eq. (\ref{Fa_el}) helps in understanding why
reducing $\mu_{c}$ also leads to an increase of $\mu_{a}/\mu_{c}$ as, under
load controlled conditions, $\mu_{a}/\mu_{c}=F_{a}/\left(  p_{m}L\mu
_{c}\right)  \propto1/(1+\mu_{c}^{2}c_{1})$ with $c_{1}>0$. The friction ratio
$\mu_{a}/\mu_{c}$ is also investigated in Figure \ref{fig_elast2}, this time
as a function of the profile mean square slope $m_{2}$, showing that, in
agreement with Refs \cite{Persson2001,menga2016visco}, increasing $m_{2}$
leads to a less than proportional increase of $\mu_{a}$. Interestingly,
assuming $m_{2}$ of order unity, as indeed measured for asphalt surfaces
\cite{Lorenz2013},\ a friction coefficient $\mu_{a}$\ due to coupling effects
as high as $10\div20\%$ of $\mu_{c}$ is predicted, thus resulting in a
non-negligible coupling effect on the overall friction force opposing the motion.

\subsubsection{Viscoelastic friction}

The physical picture drawn above for elastic materials is still valid in the
case of viscoelastic thin coatings. However, in the presence of
viscoelasticity a higher degree of contact pressure asymmetry is expected due
to delayed material response, which leads to viscous bulk energy dissipation
even in the case of semi-infinite uncoupled contacts. As in the previous case
section, since we focus on coupling arising from the finite thickness of the
viscoelastic layers involved in the contact, with referernce to the scheme
reported in Figure \ref{fig1}, we assume $\beta=0$.\begin{figure}[ptbh]
\begin{center}
\includegraphics[height=53mm]{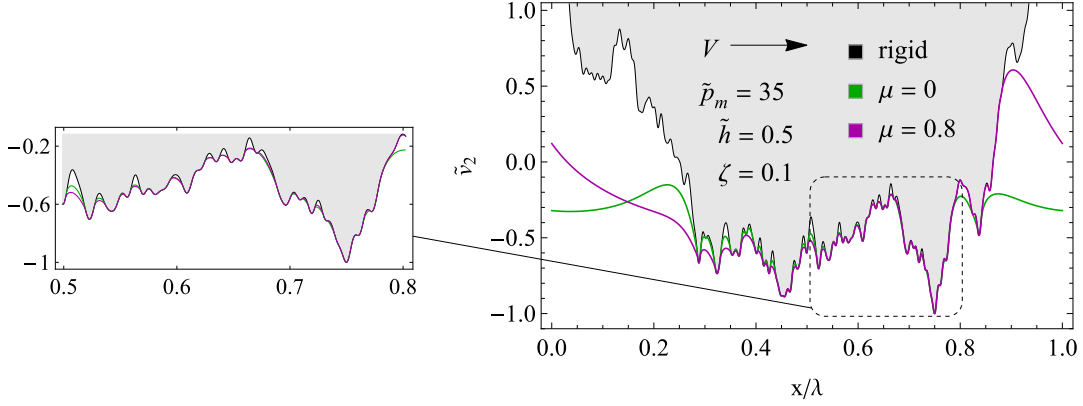}
\end{center}
\caption{The viscoelastic dimensionless normal displacement $\tilde{v}%
_{2}=\left(  v_{2}+\Delta\right)  /\Lambda-1$, under fixed normal load, in
both coupled ($\mu_{c}=0.8$) and uncoupled ($\mu_{c}=0$) conditions. Results
are for $\beta=0$ and $m_{2}=0.018$. }%
\label{fig2}%
\end{figure}

Global effect of coupling on the contact is visualized in figure \ref{fig2},
where the normal displacement field of a thin viscoelastic coating is reported
for both coupled (i.e. $\mu_{c}=0.8$) and uncoupled (i.e. $\mu_{c}=0$). A
significant degree of asymmetry between the leading (right-hand side) and
trailing (left-hand side) edges of each contact spots is observed for both
cases. Nonetheless, in coupled conditions such a behavior is even heightened.

Let us recall that the indeter slides at constant velocity $V$. Under this
condition any quantity $f\left(  x,t\right)  $ related to the contact problem
depends on space and time through the relation $f\left(  x,t\right)  =f\left(
x-Vt\right)  $. So, performing a Fourier transforms leads to $f\left(
q,\omega\right)  =\int dxdtf\left(  x-Vt\right)  e^{-i\left(  qx+\omega
t\right)  }=\delta\left(  \omega+qV\right)  f\left(  q\right)  $, and in the
case of viscoleastic coatings, Eq. (\ref{Fa}) takes the form%
\begin{equation}
F_{a}=\frac{1}{2\pi}\int\frac{\mu_{c}q^{2}\operatorname{Re}\tilde{E}\left(
qV\right)  S_{12}\left(  \left\vert q\right\vert h\right)  +q\left\vert
q\right\vert \operatorname{Im}\tilde{E}\left(  qV\right)  S_{22}\left(
\left\vert q\right\vert h\right)  }{S_{22}^{2}\left(  \left\vert q\right\vert
h\right)  +\mu^{2}S_{12}^{2}\left(  \left\vert q\right\vert h\right)
}\left\vert u_{2}\left(  q\right)  \right\vert ^{2}dq \label{Fa_visco}%
\end{equation}
where $\tilde{E}\left(  \omega=Vq\right)  =E_{0}+i\omega\tau/\left(
1+i\omega\tau\right)  E_{1}$ is the complex viscoelastic modulus.

Interestingly, in the case of uncoupled contacts of thin viscoelastic layers
(i.e. for $\mu_{c}=0$), the friction coefficient $\mu_{a,0}$, arising from the
asymmetric contact pressure caused only by viscoelastic hysteresis can be
calculated from Eqs. (\ref{mua},\ref{Fa_visco}) as%
\[
\mu_{a,0}=\frac{1}{2\pi Lp_{m}}\int q\left\vert q\right\vert \frac
{\operatorname{Im}\tilde{E}\left(  qV\right)  }{S_{22}\left(  \left\vert
q\right\vert h\right)  }\left\vert u_{2}\left(  q\right)  \right\vert ^{2}dq
\]

Building on the same arguments of Eq. (\ref{energ}), the viscoelastic layer
energy balance gives
\begin{equation}
\dot{W}=V\left[  \int_{\Omega}p\left(  x\right)  v_{2}^{\prime}\left(
x\right)  dx-\mu_{c}\int_{\Omega}p\left(  x\right)  v_{1}^{\prime}\left(
x\right)  dx\right]  \label{diss_visco}%
\end{equation}
where $\dot{W}$ is the the contribution to energy dissipation per unit time
due to the hysteretic behavior of the viscoelastic material..

\begin{figure}[ptbh]
\begin{center}
\centering\subfloat[\label{fig7a}]{\includegraphics[height=55mm] {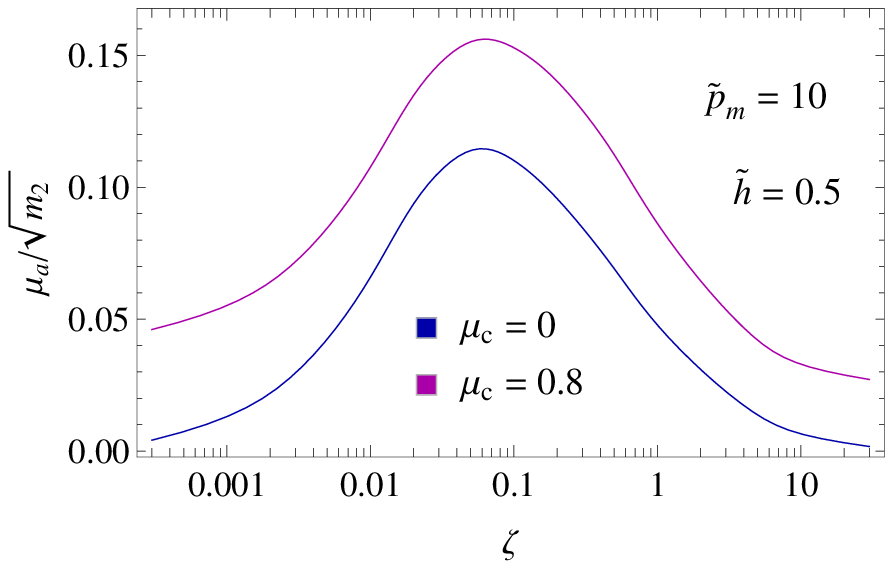}}
\hfill\subfloat[\label{fig7b}]{\includegraphics[height=55mm]{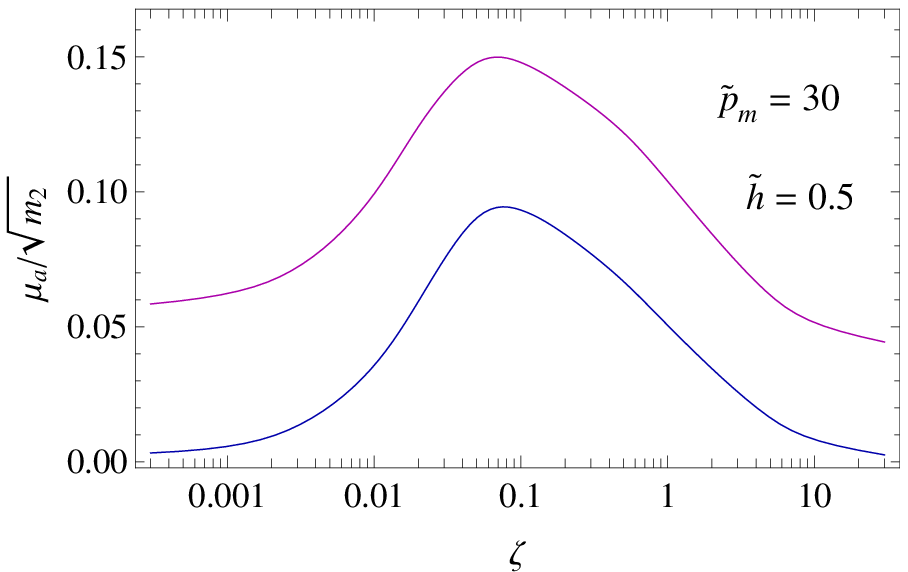}}
\end{center}
\caption{The normalized friction coefficient $\mu_{a}/\sqrt{m_{2}}$ for
viscoelastic coatings as a function of the dimensionless sliding speed $\zeta$
under different normal loads. Results refer to $\beta=0$ and $m_{2}=0.018$.}%
\label{fig7}%
\end{figure}

Figures \ref{fig7} show the friction coefficient $\mu_{a}$ due to asymmetric
contact pressure for both coupled (i.e. $\mu_{c}=0.8$) and uncoupled (i.e.
$\mu_{c}=0$) conditions as a function of the dimensionless sliding speed
$\zeta$ for two different values of the dimensionless normal load. As
expected, regardless of the coupling, curves follow the well-known bell shaped
trend with respect to the sliding velocity (i.e the excitation frequency),
although in coupled conditions globally higher friction is reported. However,
even for $\zeta\rightarrow0$ and $\zeta\rightarrow\infty$ where viscoelastic
material response is almost elastic and no viscoelastic bulk dissipation
occurs, non-vanishing friction is reported due to geometric coupling.
Interestingly, the shifting factors between coupled and uncoupled frictional
behavior (i.e. the difference between the pink and blue curves) depends on the
excitation frequency, resulting higher at lower sliding velocity. Furthermore,
comparing Figures \ref{fig7a}-\ref{fig7b}, we observe that $\mu_{a}$ in
coupled contacts appears less sensitive to the normal load variation compared
to uncoupled contact conditions.

\begin{figure}[ptbh]
\begin{center}
\centering\subfloat[\label{fig8a}]{\includegraphics[height=53mm] {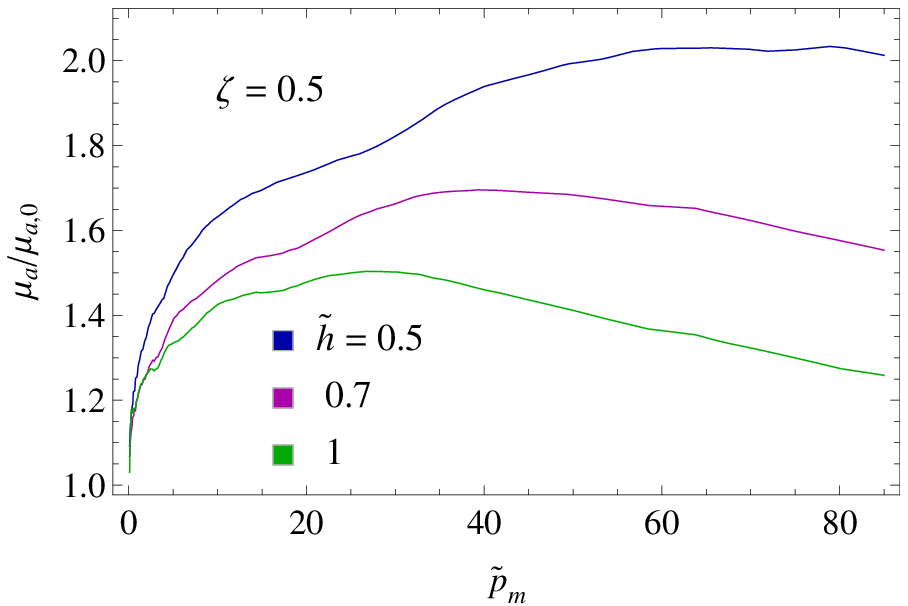}}
\hfill\subfloat[\label{fig8b}]{\includegraphics[height=52mm]{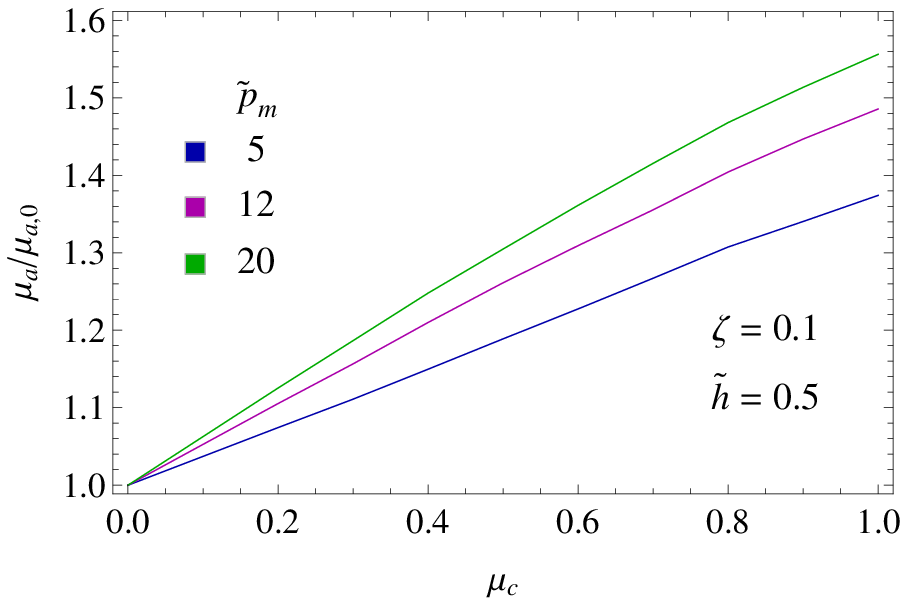}}
\end{center}
\caption{The friction ratio $\mu_{a}/\mu_{a,0}$ between the coupled and
uncoupled viscoelastic friction coefficients as a function of: (a) the
dimensionless contact mean pressure $\tilde{p}_{m}/\sqrt{m_{2}}$, for
different dimensionless coating thickness $\tilde{h}$; \ (b) the Coulomb
friction coefficient $\mu_{c}$, for different values of $\tilde{p}_{m}%
/\sqrt{m_{2}}$. Results are given for $\beta=0$ and $m_{2}=0.018$.}%
\label{fig8}%
\end{figure}

In Figures \ref{fig8} we report results in terms of the friction ratio
$\mu_{a}/\mu_{a,0}$. Specifically, in Fig. \ref{fig8a}, the effect of
$\tilde{p}_{m}$ is investigated for three different values of $\tilde{h}$. Of
course, according to Eqs. (\ref{G12}-\ref{C}) thinner coatings lead to higher
degrees of \textit{geometric} coupling. In Fig. \ref{fig8b} the effect of the
Coulomb friction coefficient $\mu_{c}$ is explored at relatively low contact
mean pressures. As already discussed for elastic contacts, increasing $\mu
_{c}$ exacerbates the effects of the normal-tangential coupling, thus leading
to\ larger values of $\mu_{a}$. Nonetheless, depending on the specific value
of the normal load, a saturation of the phenomenon is also expected.

\begin{figure}[ptbh]
\begin{center}
\includegraphics[width=0.5\textwidth]{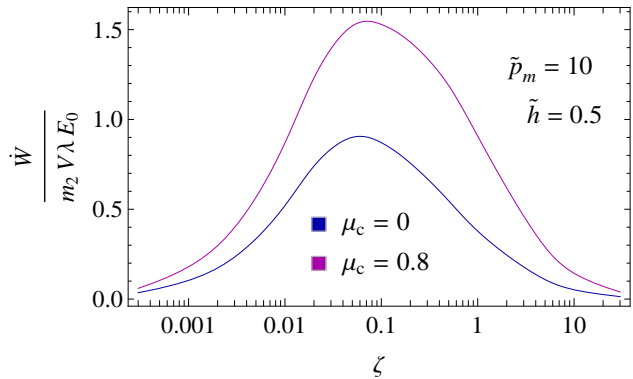}
\end{center}
\caption{The dimensionless bulk energy dissipation $\dot{W}/(m_{2}V\lambda
E_{0})$ in viscoelastic thin coatings as a function of the dimensionless
sliding speed $\zeta$. Results are given for $\beta=0$ and $m_{2}=0.018$.}%
\label{fig9}%
\end{figure}

Figure \ref{fig9} investigates the bulk energy dissipation in viscoelastic
coatings. In coupled conditions (i.e. $\mu_{c}\neq0$) increased bulk energy
dissipation is observed as, from Eq. (\ref{diss_visco}) also the tangential
deformations can contribute to the viscoelastic dissipation, depending on the
sign of the second right-hand integral. Interestingly, since the energy
dissipated is converted into bulk heat, such a peculiar behavior of
viscoelastic thin layers in the presence of interfacial friction\ may be
relevant when aiming at controlling the material warming. It is the case, for
instance, of tyres in which a key component is the tread (a thin coating on
the underlaying stiffer tyre structure) whose warming is crucial for the
overall system performance. Neglecting the effect of geometric coupling due to
the layer thickness in such systems may significantly alter the theoretical predictions.

\subsection{Viscoelastic contact behavior}

In this section we investigated the rough contact results is terms of mean
contact quantities (contact area, mean pressure and mean penetration) and
displacement fields. We refer our analysis to the case of thin viscoelastic
coatings shown in Figure \ref{fig1}, assuming $\beta=0$, so that the only
source of coupling is related to the layer thickness. Notably, in Ref.
\cite{menga2019geom} a similar investigation has been devoted to the case of
elastic thin layers.

\subsubsection{Mean contact quantities}

\begin{figure}[ptbh]
\begin{center}
\centering\subfloat[\label{fig3a}]{\includegraphics[height=54mm] {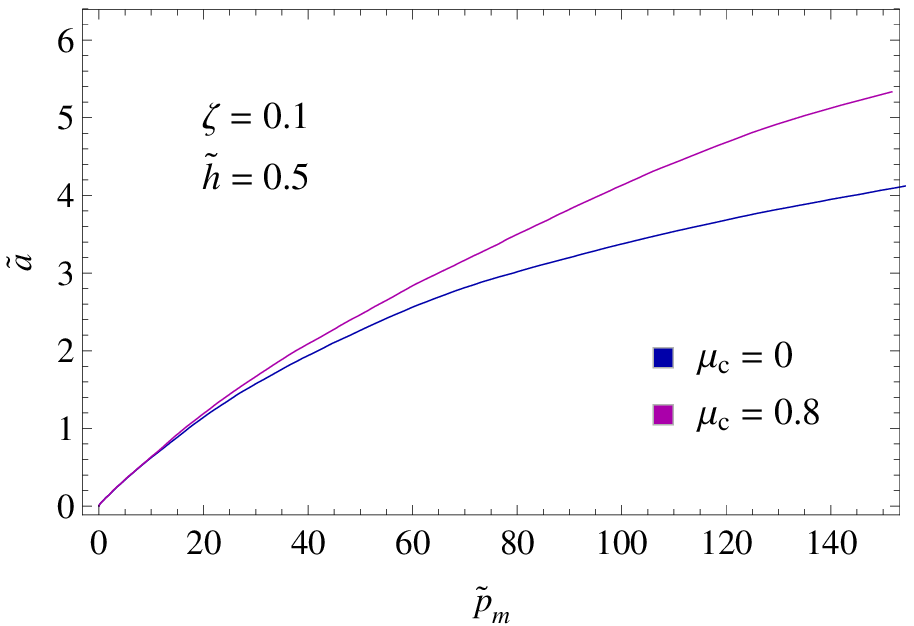}}
\hfill\subfloat[\label{fig3b}]{\includegraphics[height=54mm]{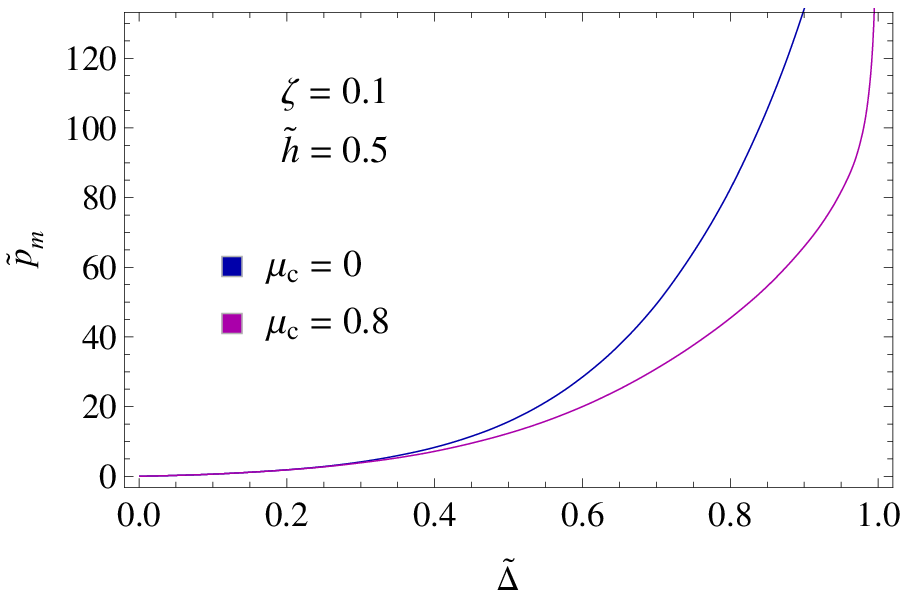}}
\end{center}
\caption{The dimensionless contact area $\tilde{a}$ as a fucntion of the
dimensionless contact mean pressure $\tilde{p}_{m}$ (a); and the dimensionless
contact mean pressure $\tilde{p}_{m}$ as a function of the dimensionless
contact penetration $\tilde{\Delta}$. Full contact is for $\tilde{a}=2\pi.$
Results refer to $\beta=0$ and $m_{2}=0.018$.}%
\label{fig3}%
\end{figure}

Figures \ref{fig3} present the main contact behavior of the viscoelastic thin
layer in terms of the main contact quantities. Interestingly, regardless of
the physical quantity under investigation, we note that the difference among
the coupled (i.e. $\mu_{c}\neq0$) and uncoupled (i.e. $\mu_{c}=0$) results
increases with $\tilde{p}_{m}$ increasing. This can be explained, in agreement
with Ref. \cite{menga2019geom}, by observing that, from Eq. (\ref{frict}), the
shear stresses are proportional to the normal pressure, thus, at low values of
$\tilde{p}_{m}$, even in the case of $\mu_{c}\neq0$ low tangential stresses
occurs thus leading to a poor normal-tangential coupling. On the contrary,
increasing $\tilde{p}_{m}$ leads to higher shear stresses and, in turn, to
larger coupling effects.

\begin{figure}[ptbh]
\begin{center}
\centering\subfloat[\label{fig4a}]{\includegraphics[height=53mm] {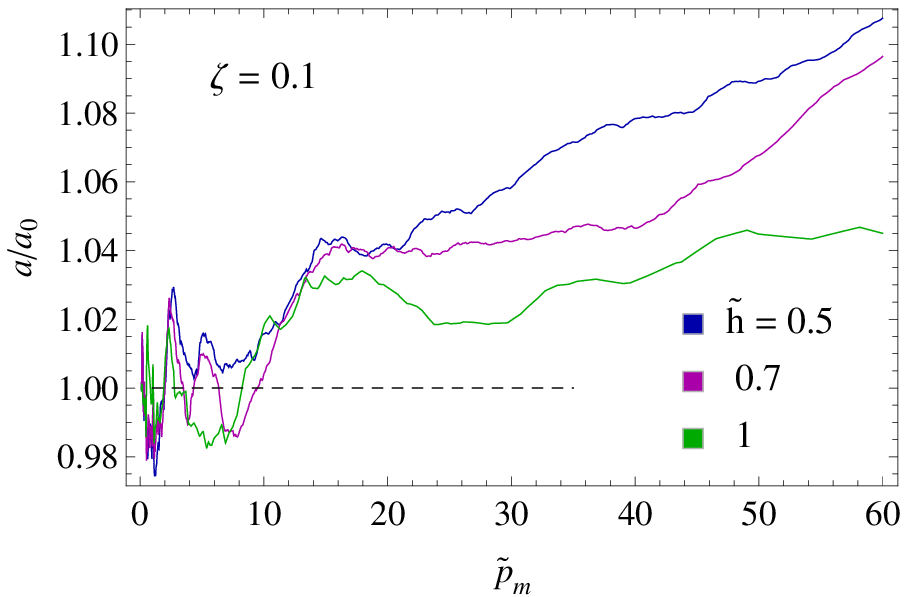}}
\hfill\subfloat[\label{fig4b}]{\includegraphics[height=56mm]{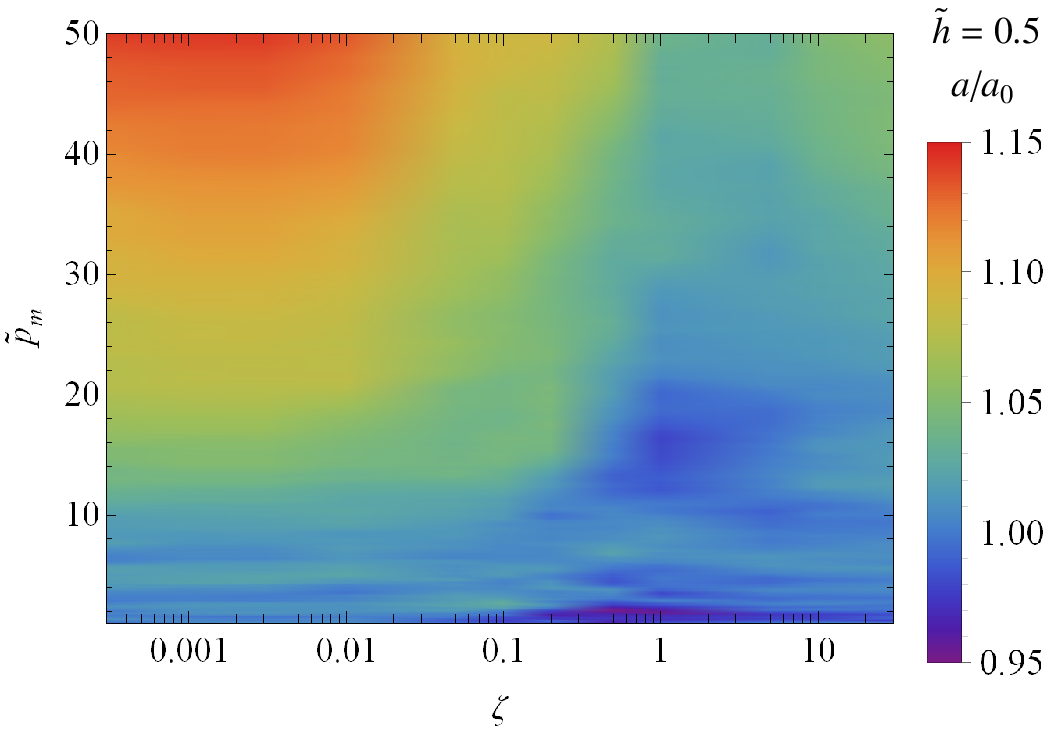}}
\end{center}
\caption{The contact size ratio $a/a_{0},$ between the contact are length
associated to coupled ($\mu_{c}=0.8$) and uncoupled ($\mu_{c}=0$) conditions,
as a fucntion of: (a) the dimensionless contact mean pressure $\tilde{p}_{m}$,
for different values of the dimensionless layer thickness $\tilde{h}$; (b) the
dimensionless contact mean pressure $\tilde{p}_{m}$ and the dimensionless
sliding speed $\zeta.$ Results refer to $\beta=0$ and $m_{2}=0.018.$}%
\label{fig4}%
\end{figure}

Figure \ref{fig4a} shows the contact size ratio $a/a_{0}$ as a function of the
contact mean pressure $\tilde{p}_{m}$, for different values of the
dimensionless coating thickness $\tilde{h}$, where $a$ represent the contact
size in coupled conditions, and $a_{0}$ refers to the frictionless uncoupled
ones. Notably, the initial results scattering depends on the fact that, at low
values of $\tilde{p}_{m}$, the contact occurs on very small contact spots,
mostly localized on top of the roughness crests. This dramatically affects the
statistical sampling of the rough profiles, thus leading to noisy results.
However, we can still observe that, for $\tilde{p}_{m}\lessapprox10$, a slight
coupling effect is reported on the final contact area size. On the contrary,
for $\tilde{p}_{m}\gtrapprox10$, since the contact size globally increases,
frictional shear stress start to play a key role on the contact behavior.
Indeed, according to Eqs. (\ref{10},\ref{G11}-\ref{G22}), due to
\textit{geometric} coupling, the shear stresses also affect the normal
displacements of the contacting interfaces, thus leading to a marked increase
of the contact area in coupled conditions compared to the frictionless case
where no coupling occurs. As expected, increasing the coating thickness
flattens the ratio $a/a_{0}$ towards the unity value, as with $\beta=0$ Eqs.
(\ref{G12},\ref{C}) show that coupling terms monotonically decrease with
$\tilde{h}$ increasing, eventually leading, for $\tilde{h}\gtrapprox10$, to
the uncoupled half-space contact behavior (see also Ref. \cite{menga2019geom}).

Figure \ref{fig4b} shows a contour map of the contact size ratio $a/a_{0}$,
for a specific value of $\tilde{h}$, as a function of both the dimensionless
sliding speed $\zeta$ and $\tilde{p}_{m}$. Interestingly, we observe that
stronger coupling effects are predicted at low sliding speed, as the ratio
$a/a_{0}$ at low values of $\zeta$ is globally higher than what observed at
high values of $\zeta$. However, the trend of $a/a_{0}$\ vs $\zeta$ is
non-monotonic, presenting a minimum at $\zeta\approx1$. The results shown in
Fig. \ref{fig4b} offer interesting perspectives, as a contact length increase
as high as $15\%$ can be achieved in coupled conditions. Such a large
difference suggests that in real life contact problems involving sliding thin
viscoelastic layers (e.g. the accurate detection of finger prints on touch
screens), the real contact area can be significantly underestimated by
neglecting normal-tangential coupling effects. Moreover, according to Eqs.
(\ref{G12},\ref{C}), the thinner the viscoelastic layer involved, the larger
the error in the contact area predicted by uncoupled models.

\subsubsection{Normal displacement field}

A detailed comparison of the normal displacement fields reported in Figure
\ref{fig2} reveals that coupled and uncoupled contacts presents macroscopic
differences in the deformed shape of the viscoelastic layer. However, the
close-up shown in the same Figure suggests that most of the differences
actually occur to the larger spatial scales, as the behavior at the smaller
scales appears much more similar among the two cases

\begin{figure}[ptbh]
\begin{center}
\centering\subfloat[\label{fig5a}]{\includegraphics[height=55mm] {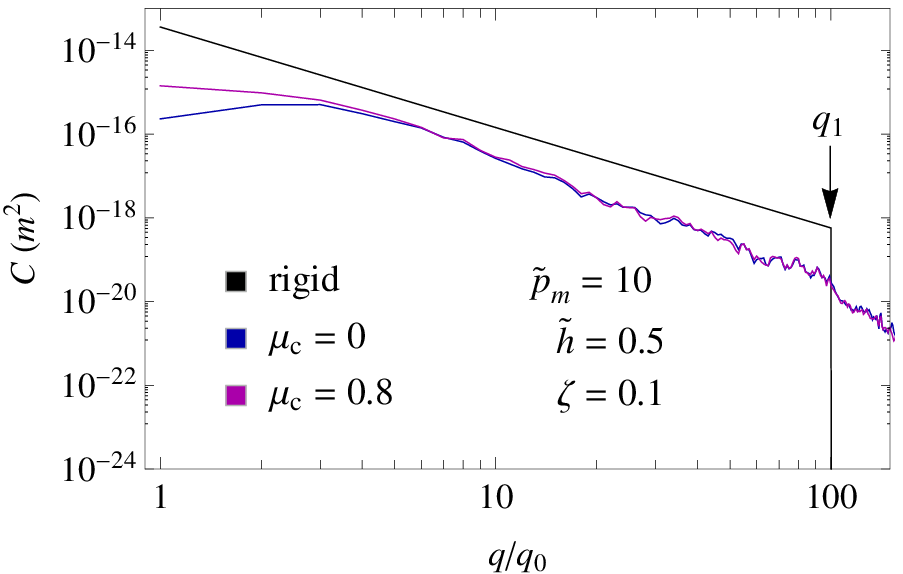}}
\hfill\subfloat[\label{fig5b}]{\includegraphics[height=55mm]{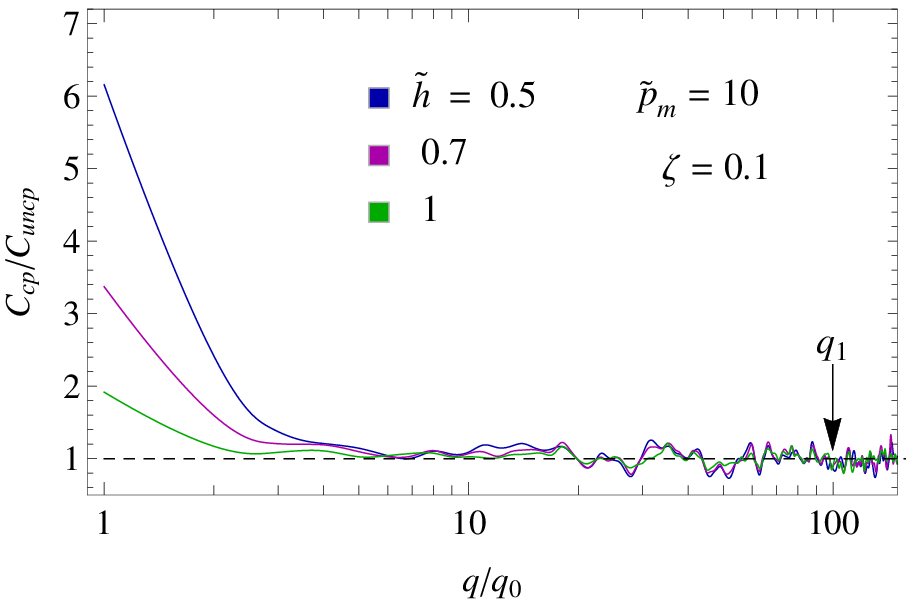}}
\end{center}
\caption{(a) the PSD $C$ of the normal displacement fields for both the
coupled and uncoupled conditions as a function of the dimensionless spatial
frequency $q/q_{0}$; the ratio $C_{cp}/C_{uncp}$ of the mormal displacement
PSD under coupled and uncoupled conditions as a function of the dimensionless
spatial frequency $q/q_{0}$, for different coating dimensionless thickness
$\tilde{h}.$ Coupled conditions refer to $\mu_{c}=0.8$, whereas the uncoupled
case is for $\mu_{c}=0$. Results refer to $\beta=0$ and $m_{2}=0.018$.}%
\label{fig5}%
\end{figure}

In this regard, Figure \ref{fig5a} offers a quantitative comparison between
the Power Spectral Density (PSD) $C\left(  q\right)  $ of the deformed
profiles under coupled (i.e. $\mu_{c}\neq0$) and uncoupled (i.e. $\mu_{c}=0$)
conditions, under load controlled conditions. The rigid rough profile PSD is
also shown (black curve) to help the comparison. Results indicate that in the
presence of geometric coupling enhanced large scale deformations occur (i.e.
at lower spatial frequencies $q$) compared to uncoupled case. Such a result is
even more clearly shown in Fig. \ref{fig5b}, where we plot the ratio
$C_{cp}/C_{uncp}$ between the deformed profile PSD $C_{cp}$ under coupled and
$C_{uncp}$ uncoupled conditions. According to Eqs. (\ref{G12},\ref{C}), the
coupling effect on each deformation scale $\lambda_{i}=2\pi/q$ increases with
$qh$ reducing, thus leading to larger difference between coupled and uncoupled
response at the larger scales. On the contrary, at smaller scales (i.e. for
$\lambda_{i}=2\pi/q\ll h$) the contact behavior recovers the one expected in
the case of half-space contacts, as the normal-tangential coupling terms of
Eqs. (\ref{G12},\ref{C}) vanish. Building on dimensional arguments, we expect
that for $q/q_{0}>\rho/\tilde{h}$ the coupling effect should vanish (with
$\rho$ being a constant of order unity), as indeed reported in Fig.
\ref{fig5b} where we observe that the spatial frequency $q$ at which the
coupling effects vanish (i.e. $C_{cp}/C_{uncp}\approx1$) increases with
$\tilde{h}$ decreasing.

\begin{figure}[ptbh]
\begin{center}
\centering\subfloat[\label{fig6a}]{\includegraphics[height=50mm] {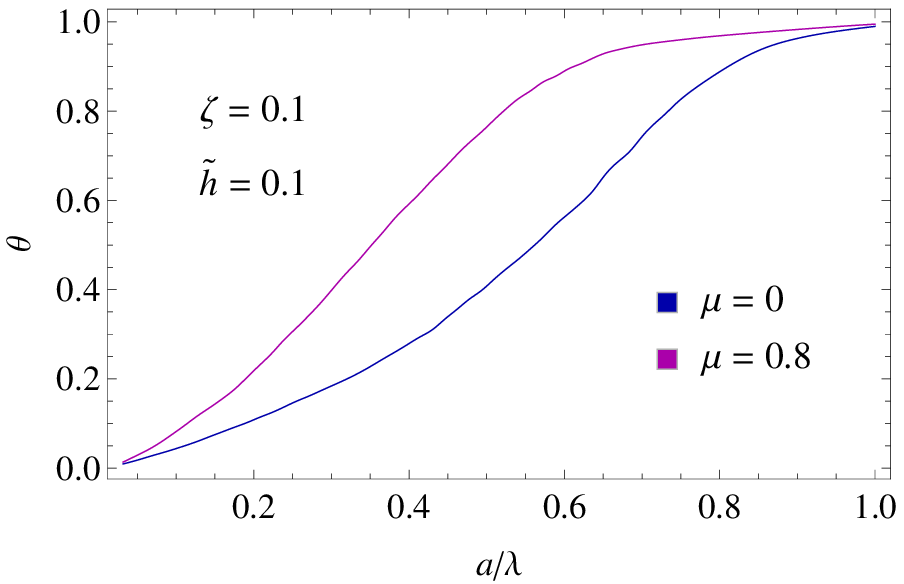}}
\hfill\subfloat[\label{fig6b}]{\includegraphics[height=59mm]{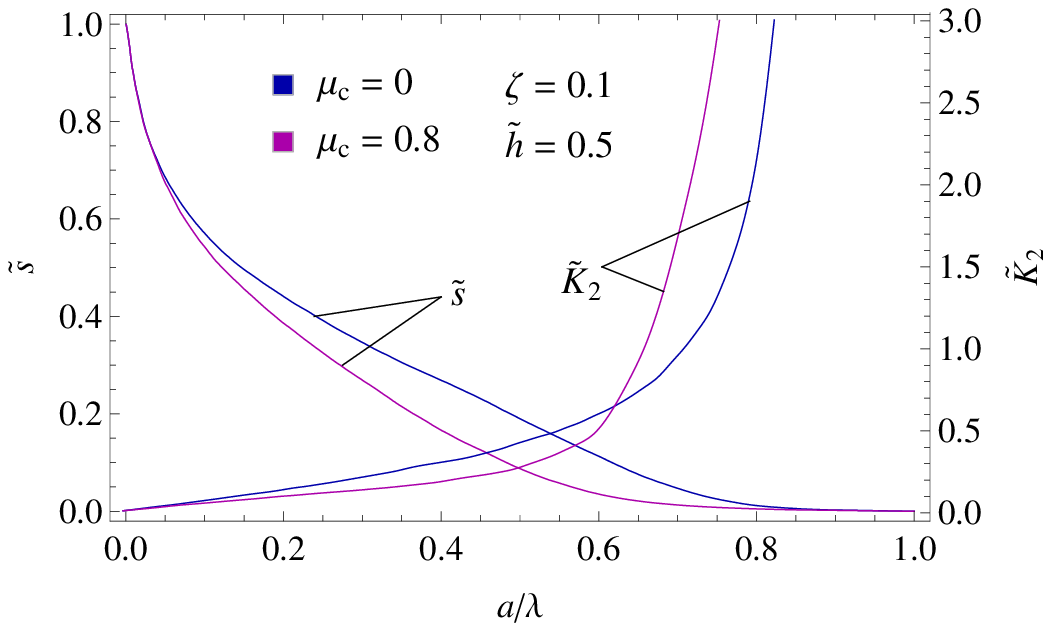}}
\end{center}
\caption{The ratio $\theta$ between the frictional coupled and frictionless
uncoupled systems mean square heights of the deformed profiles (a), and the
dimensionless contact mean separation $\tilde{s}=s/\Lambda$ and the
dimensionless normal stiffness $\tilde{K}_{2}=10^{-3}d\tilde{p}_{m}%
/d\tilde{\Delta}$ as a function of the contact area fraction $a/\lambda$. For
the frictional coupled case, the friction coefficient is $\mu_{c}=0.8.$}%
\label{fig6}%
\end{figure}

The effect of the \textit{geometric} coupling on the spectrum of the
viscoelastic normal displacement field can be further explored by defining
$\theta$ as the ratio between the mean square roughness of the deformed
profile and the rigid one, i.e.%
\begin{equation}
\theta=\frac{\int_{q_{0}}^{q_{1}}C\left(  q\right)  dq}{\int_{q_{0}}^{q_{1}%
}C_{r}\left(  q\right)  dq}=\frac{\left\langle u_{2}^{2}\right\rangle
}{\left\langle r^{2}\right\rangle }%
\end{equation}
where $\left\langle u_{2}^{2}\right\rangle $ and $\left\langle r^{2}%
\right\rangle =r_{rms}^{2}$ are the mean square roughness of the deformed
profile and the rigid indenter, respectively.

In Figure \ref{fig6a} we show $\theta$ as a function of the contact area
fraction $a/\lambda$ (notably, $\lambda$ is the full contact area value) for
both the coupled contact condition and the uncoupled one. Of course, in both
cases, for $a/\lambda\rightarrow1$ the value of $\theta$ tends to unity as,
regardless of the degree of coupling, the coating displacements completely
match the rigid rough profile. However, for $a/\lambda<1$, higher values of
$\theta$ are reported in coupled conditions, thus indicating that in the
presence of coupling (i.e. for $\mu_{c}\neq0$) the deformed interface is
globally closer to the rigid counterpart than in uncoupled conditions.
Although the contact spots can be differently located in each case, the
comparison of Fig. \ref{fig6a} is performed at given total contact length thus
the difference in $\theta$ has to be mostly ascribed to the non-contact
region, where the displacements of the deformable coating in coupled
conditions present thinner gaps to the rigid profile compared to the uncoupled
case. This can be further investigated by looking at the contact mean
separation%
\[
s=\frac{1}{\lambda}\int_{\lambda}g\left(  x\right)  dx=\Lambda-\Delta
\]
where $g\left(  x\right)  =r\left(  x\right)  -\left[  u\left(  x\right)
+\Delta-\Lambda\right]  $ is the gap function between the deformed
viscoelastic surface and the rigid profile (notably, $g\left(  x\right)  =0$
for $x\in\Omega$). Indeed, in Fig. \ref{fig6b} we plot the dimensionless
normal separation $\tilde{s}$ against $a/\lambda$, showing that thinner gaps
are expected in the case of frictional coupled contacts compared to the
frictionless uncoupled case. In the same figure, also the contact stiffness
$K_{2}=dp_{m}/d\Delta$ is shown indicating that the additional normal
deformation introduced by the coupling term $G_{21}$ in Eq. (\ref{10}) leads
to different results, depending on the contact area size. Indeed, for the case
under investigation, for a\ contact area length up to 60\% of the full contact
length $\lambda$ (i.e. in most of the practical applications) the frictional
coupled case present lower contact stiffness compared to the frictionless
uncoupled one. Only at very large contact area the scenario is reversed.

\section{Conclusions}

In this work we have investigated the frictional behavior of thin coatings
bonded to rigid substrates in sliding rough contact. The analysis aims at
exploring the effect of the peculiar coupling between the normal and
tangential displacement fields arising in the case of thin bodies, even for
similar contacting materials. The presence of Coulomb friction interactions,
through non-null interfacial tangential stresses, activate the coupling
effects, which instead vanishes in frictionless contacts.

We found that, in the presence of sufficiently thin layers, normal-tangential
coupling occurs so that even in purely elastic rough contacts, where no bulk
dissipation occurs, asymmetric contact pressure distributions are reported
during sliding, thus resulting in a tangential force opposing the motion
larger than that resulting from Coulomb friction considered in isolation. The
friction force increase is governed by the mean square slope of the rough
counterpart, resulting non-negligible in the range of slope values typical of
real surfaces. A similar behavior is observed also in the case of thin
viscoelastic layers, where in the presence of interfacial Coulomb friction
normal-tangential coupling occurs and higher viscoelastic friction is achieved
compared to the uncoupled case, this time also associated to higher bulk
energy dissipation. As a consequence, since in real contacts of thin
viscoelastic layers interfacial friction is more likely to occur, higher
overall friction and faster bulk warming can be expected, with non-negligible
effect on the tribological behavior of the interface (e.g. tyre frictional performance).

Results show that the \textit{geometric} coupling between normal-tangential
fields enhances the normal displacements on large scales. This, under given
normal load, corresponds to significantly larger contact areas size in coupled
conditions compared to uncoupled systems, and a resulting lower normal
stiffness of the contact interface.

This study proves that, in contacts involving thin deformable coatings bonded
to significantly stiffer substrates, neglecting the effect of cross-coupled
interfacial shear stresses may lead to significant underestimation of the
overall friction and contact area.

\begin{acknowledgement}
This project has received funding from the European Union's Horizon 2020
research and innovation programme under the Marie Sk\l odowska- Curie grant
agreement no. 845756 (N.M. Individual Fellowship). D.D. acknowledges the
support received from the Engineering and Physical Science Research Council
(EPSRC) through his Established Career Fellowship EP/N025954/1. This work was
partly supported by the Italian Ministry of Education, University and Research
under the Programme \textquotedblleft Progetti di Rilevante Interesse
Nazionale (PRIN)\textquotedblright, Grant Protocol 2017948, Title: Foam
Airless Spoked Tire -- FASTire (G.C.)
\end{acknowledgement}

\appendix{}

\section{Effect of dissimilar materials in thin elastic layers contacts}

The case of dissimilar elastic contacting materials can be investigated by
assuming the Dundurs' second constant $\beta\neq0$ \cite{HillsBook,BarberBook}%
. Since, for simplicity, here we focus on the contact between a deformable
solid and a rigid one, it takes the form $\beta=(1-2\nu)/2(1-\nu)\neq0$. Under
these conditions, for thin elastic layers in sliding frictional contacts, both
the \textit{material }(i.e. the first right-hand side term, vanishing for
$\beta=0$) and \textit{geometric} coupling (i.e. the first right-hand side
term) terms in Eq. (\ref{G12}) are non-vanishing, thus resulting in asymmetric
contact pressure distribution. Interestingly, the two terms have opposite
effects on the normal displacements so that, through Eqs. (\ref{Fa},\ref{Ft}),
the former term is expected to reduce the friction force, whereas the latter
term leads to globally higher friction force. Moreover, according to Ref.
\cite{menga2019geom}, dimensional arguments suggest that the correlation
length of the \textit{geometric} term (of order unity of $h$) is larger than
that associated to the \textit{material} one.

\begin{figure}[ptbh]
\begin{center}
\centering\subfloat[\label{fig_elast1}]{\includegraphics[height=49mm] {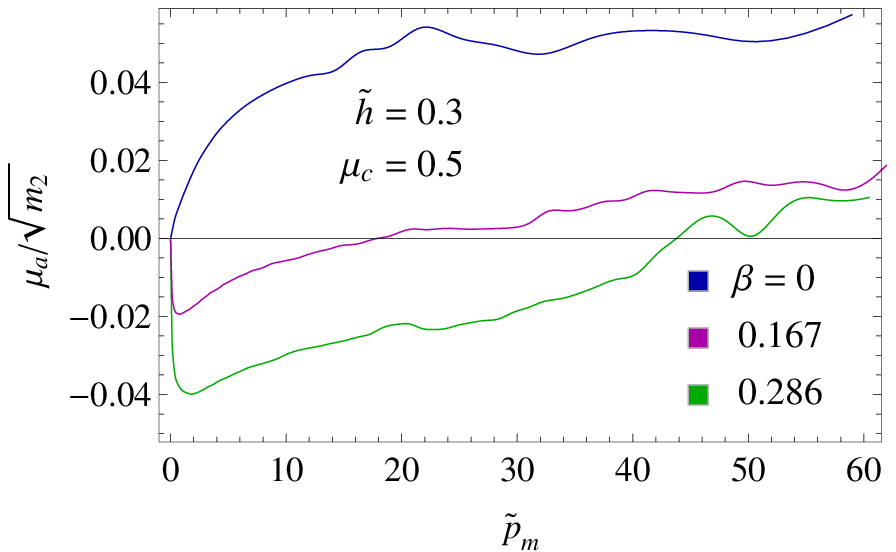}}
\hfill
\subfloat[\label{fig_elast2}]{\includegraphics[height=49mm]{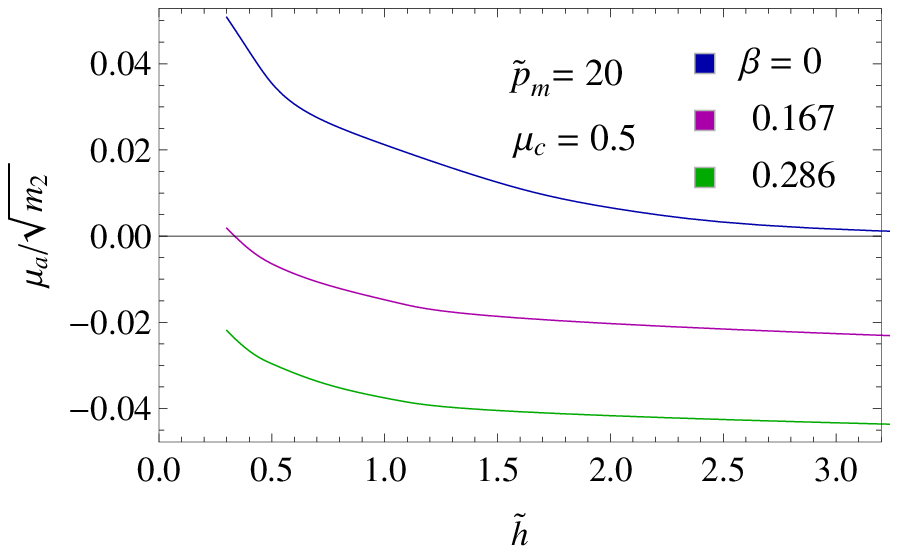}}
\end{center}
\caption{The friction coefficient $\mu_{a}$ due to asymmetric contact pressure
distribution as a function of (a) the dimensionless contact mean pressure
$\tilde{p}_{m}$ and (b) the elastic layer dimensionless thickness $\tilde{h}$.
Results results refers to $m_{2}=0.018$.}%
\end{figure}

Figures \ref{fig_elast} show the friction coefficient $\mu_{a}$ resulting from
asymmetric contact pressure in the case of elastic contacts, for different
values of $\beta$. Specifically, in Figure \ref{fig_elast1} $\mu_{a}$ is shown
against $\tilde{p}_{m}$. For $\beta=0.28$ (i.e. for $\nu=0.3$), at very low
contact pressure, since the contact spots are sufficiently small, the material
coupling effect is enhanced due to its shorter correlation length, thus
leading to a reduction of the total friction force $F_{t}$ opposing the
relative layer-indenter motion (see Eq. (\ref{Ft})). However, as the normal
load increases, larger contact spots are experienced and the
\textit{geometric} coupling starts playing a key role, thus increasing
$\mu_{a}$. For sufficiently large values of $\tilde{p}_{m}$, the geometric
coupling term is dominant, thus $\mu_{a}>0$ and overall increase of the total
friction is expected compared to the uncoupled case (i.e. $\mu_{c}=0$).
Furthermore, figure \ref{fig_elast2} shows the effect of the elastic coating
thickness on $\mu_{a}$. Of course, only the \textit{geometric} term is
affected by $h$, thus curved with $\beta=0$ and $\beta\neq0$ are shifted by a
quantity which can be roughly estimated as proportional to the
\textit{material} term. Notably, for $h\rightarrow0$ in both cases we expect
$\mu_{a}\rightarrow0$, as $u\left(  q\right)  \rightarrow0$ in the second of
Eqs. (\ref{Fa_el}).


\begin{thebibliography}{99}                                                                                               %


\bibitem {Dahotre2005}Dahotre, N. B., \& Nayak, S. (2005). Nanocoatings for
engine application. Surface and Coatings Technology, 194(1), 58-67.

\bibitem {Kano2006}Kano, M. (2006). Super low friction of DLC applied to
engine cam follower lubricated with ester-containing oil. Tribology
International, 39(12), 1682-1685.

\bibitem {Voigt2012}Voigt, D., Karguth, A., \& Gorb, S. (2012). Shoe soles for
the gripping robot: Searching for polymer-based materials maximising friction.
Robotics and Autonomous Systems, 60(8), 1046-1055.

\bibitem {GW1966}Greenwood, J. A., \& Williamson, J. P. (1966). Contact of
nominally flat surfaces. Proc. R. Soc. Lond. A, 295(1442), 300-319.

\bibitem {BGT}Bush, A.W., Gibson, R.D., Thomas, T.R., TheElastic Contact of a
Rough Surface, Wear 1975;35:87-111.

\bibitem {Persson2001}Persson B.N.J.,Theory of rubber friction and contact
mechanics, Journal of Chemical Physics 2001;115:3840 -3861.

\bibitem {YangPersson2008}Yang, C. and Persson, B.N.J., Molecular Dynamics
Study of Contact Mechanics: Contact Area and Interfacial Separation from Small
to Full Contact, Phis, Rev. Lett. 2008;100.

\bibitem {Menga2014}Menga, N., Putignano, C., Carbone, G., \& Demelio, G. P.
(2014). The sliding contact of a rigid wavy surface with a viscoelastic
half-space. Proc. R. Soc. A, 470(2169), 20140392.

\bibitem {menga2018}Menga, N., Carbone, G., \& Dini, D. Do uniform tangential
interfacial stresses enhance adhesion? Journal of the Mechanics and Physics of
Solids 2018;112:145-156.

\bibitem {menga2018corr}Menga, N., Carbone, G., \& Dini, D. (2019).
Corrigendum to:\textquotedblleft Do uniform tangential interfacial stresses
enhance adhesion?\textquotedblright. Journal of the Mechanics and Physics of
Solids 2019;133:103744. Doi: https://doi.org/10.1016/j.jmps.2019.103744

\bibitem {menga2018bis}Menga, N., \& Carbone, G. (2019). The surface
displacements of an elastic half-space subjected to uniform tangential
tractions applied on a circular area. European Journal of Mechanics-A/Solids,
73, 137-143.

\bibitem {Hyun2004}Hyun, S., Pei, L., Molinari, J.-F., Robbins, M.O.,
Finite-element analysis of contact between elastic self-affine surfaces, Phys.
Rev. E 2004;70.

\bibitem {Campana2008}Campana C., Mueser M.H. and Robbins M.O., Elastic
contact between self-affine surfaces: comparison of numerical stress and
contact correlation functions with analytic predictions. J. Phys. Condens.
Matter 2008;20(35).

\bibitem {Pastewka2016}Pastewka, L., \& Robbins, M. O. Contact area of rough
spheres: Large scale simulations and simple scaling laws. Applied Physics
Letters 2016;108(22):221601.

\bibitem {DiniMedina}Medina S. and Dini D., A numerical model for the
deterministic analysis of adhesive rough contacts down to the nano-scale.
International Journal of Solids and Structures 2014;51(14):2620-2632.

\bibitem {Muser2017}M\"{u}ser, M. H., Dapp, W. B., Bugnicourt, R., Sainsot,
P., Lesaffre, N., Lubrecht, T. A., ... \& Rohde, S. (2017). Meeting the
contact-mechanics challenge. Tribology Letters, 65(4), 118.

\bibitem {mengaVpeeling}Menga, N., Afferrante, L., Pugno, N. M., \& Carbone,
G. (2018). The multiple V-shaped double peeling of elastic thin films from
elastic soft substrates. Journal of the Mechanics and Physics of Solids, 113, 56-64.

\bibitem {Homola1990}Homola, A. M., Israelachvili, J. N., McGuiggan, P. M., \&
Gee, M. L. (1990). Fundamental experimental studies in tribology: the
transition from \textquotedblleft interfacial\textquotedblright\ friction of
undamaged molecularly smooth surfaces to \textquotedblleft
normal\textquotedblright\ friction with wear. Wear, 136(1), 65-83.

\bibitem {Chateauminois2008}Chateauminois, A., \& Fretigny, C. (2008). Local
friction at a sliding interface between an elastomer and a rigid spherical
probe. The European Physical Journal E: Soft Matter and Biological Physics,
27(2), 221-227.

\bibitem {new2}Krick, B. A., Vail, J. R., Persson, B. N., \& Sawyer, W. G.
(2012). Optical in situ micro tribometer for analysis of real contact area for
contact mechanics, adhesion, and sliding experiments. Tribology Letters,
45(1), 185-194.

\bibitem {Fineberg2010}Ben-David, O., Cohen, G., \& Fineberg, J. (2010). The
dynamics of the onset of frictional slip. Science, 330(6001), 211-214.

\bibitem {Carbone2008}Carbone, G., \& Mangialardi, L. (2008). Analysis of the
adhesive contact of confined layers by using a Green's function approach.
Journal of the Mechanics and Physics of Solids, 56(2), 684-706.

\bibitem {Putignano2015}Putignano, C., Carbone, G., \& Dini, D. (2015).
Mechanics of rough contacts in elastic and viscoelastic thin layers.
International Journal of Solids and Structures, 69, 507-517.

\bibitem {menga2016}Menga, N., L. Afferrante, and G. Carbone. Adhesive and
adhesiveless contact mechanics of elastic layers on slightly wavy rigid
substrates. International Journal of Solids and Structures 2016;88:101-109.

\bibitem {menga2016visco}Menga, N., Afferrante, L. and Carbone, G., Effect of
thickness and boundary conditions on the behavior of viscoelastic layers in
sliding contact with wavy profiles, The Journal of the Mechanics and Physics
of Solids 2016;95: 517-529.

\bibitem {mengaRLRB}Menga, N., Foti, D., \& Carbone, G. Viscoelastic
frictional properties of rubber-layer roller bearings (RLRB) seismic
isolators. Meccanica 2017;52(11-12):2807-2817.

\bibitem {mengaVpeelingthin}Menga, N., Dini, D., \& Carbone, G. (2020). Tuning
the periodic V-peeling behavior of elastic tapes applied to thin compliant
substrates. International Journal of Mechanical Sciences, 170, 105331.

\bibitem {menga2019geom}Menga, N. (2019). Rough frictional contact of elastic
thin layers: The effect of geometric coupling. International Journal of Solids
and Structures, 164, 212-220.

\bibitem {Bentall1968}Bentall, R. H., \& Johnson, K. L. (1968). An elastic
strip in plane rolling contact. International Journal of Mechanical Sciences,
10(8), 637-663.

\bibitem {Nowell1988b}Nowell, D., \& Hills, D. A. (1988). Contact problems
incorporating elastic layers. International Journal of Solids and Structures,
24(1), 105-115.

\bibitem {Nowell1988c}Nowell, D., and D. A. Hills. "Tractive rolling of tyred
cylinders." International journal of mechanical sciences 30.12 (1988): 945-957.

\bibitem {HillsBook}Sackfield, A., Hills, D. A., \& Nowell, D. (2013).
Mechanics of elastic contacts. Elsevier.

\bibitem {BarberBook}Barber, J. R. (2018). Contact mechanics (Vol. 250). Springer.

\bibitem {Nowell1988}Nowell, D., Hills, D. A., \& Sackfield, A. (1988).
Contact of dissimilar elastic cylinders under normal and tangential loading.
Journal of the Mechanics and Physics of Solids, 36(1), 59-75.

\bibitem {Chen2008}Chen, W. W., \& Wang, Q. J. (2008). A numerical model for
the point contact of dissimilar materials considering tangential tractions.
Mechanics of Materials, 40(11), 936-948.

\bibitem {Chen2009}Chen, W. W., \& Wang, Q. J. (2009). A numerical static
friction model for spherical contacts of rough surfaces, influence of load,
material, and roughness. Journal of Tribology, 131(2), 021402.

\bibitem {Wang2010}Wang, Z. J., Wang, W. Z., Wang, H., Zhu, D., \& Hu, Y. Z.
(2010). Partial slip contact analysis on three-dimensional elastic layered
half space. Journal of Tribology, 132(2), 021403.

\bibitem {Elloumi2010}Elloumi, R., Kallel-Kamoun, I., \& El-Borgi, S. (2010).
A fully coupled partial slip contact problem in a graded half-plane. Mechanics
of Materials, 42(4), 417-428.

\bibitem {Jaffar1993}Jaffar, M. J. (1993). Determination of surface
deformation of a bonded elastic layer indented by a rigid cylinder using the
Chebyshev series method. Wear, 170(2), 291-294.

\bibitem {Jaffar1997}Jaffar, M. J. (1997). A numerical investigation of the
sinusoidal model for elastic layers in line contact. International journal of
mechanical sciences, 39(5), 497-506.

\bibitem {Kogut2003}Kogut, L., \& Komvopoulos, K. (2003). Electrical contact
resistance theory for conductive rough surfaces. Journal of Applied Physics,
94(5), 3153-3162.

\bibitem {MengaCiava}Menga, Nicola, and Michele Ciavarella. "A Winkler
solution for the axisymmetric Hertzian contact problem with wear and finite
element method comparison." The Journal of Strain Analysis for Engineering
Design 50.3 (2015): 156-162.

\bibitem {carb-mang-2008}Carbone, G., \& Mangialardi, L. (2008). Analysis of
the adhesive contact of confined layers by using a Green's function approach.
Journal of the Mechanics and Physics of Solids, 56(2), 684-706.

\bibitem {menga2018visco}Menga, N., Afferrante, L., Demelio, G. P., \&
Carbone, G. Rough contact of sliding viscoelastic layers: numerical
calculations and theoretical predictions. Tribology International, 2018; 122:67-75.

\bibitem {maugis}Maugis, D., Contact Adhesion and Rupture of Elastic Solids.
Springer-Verlag Berlin Heidelberg, 2000.

\bibitem {Christensen}Christensen, R., (1982). Theory of Viscoelasticity.
Academic Press

\bibitem {Persson2004}Persson, B. N., Albohr, O., Tartaglino, U., Volokitin,
A. I., \& Tosatti, E. (2004). On the nature of surface roughness with
application to contact mechanics, sealing, rubber friction and adhesion.
Journal of physics: Condensed matter, 17(1), R1.

\bibitem {Lorenz2013}Lorenz, B., Persson, B. N. J., Fortunato, G.,
Giustiniano, M., \& Baldoni, F. (2013). Rubber friction for tire tread
compound on road surfaces. Journal of Physics: Condensed Matter, 25(9), 095007.
\end{thebibliography}
\end{document}